\documentstyle[12pt,aaspp4]{article}

\def\gtsim{\ {\raise-0.5ex\hbox{$\buildrel>\over\sim$}}\ }
\def\ltsim{\ {\raise-0.5ex\hbox{$\buildrel<\over\sim$}}\ }
\slugcomment{Submitted to The Astrophysical Journal}

\begin{document}

\title{
Is the Large Magellanic Cloud Microlensing Due to an Intervening Dwarf Galaxy?
}

\author {C.  Alcock$^{1,2}$, R.A. Allsman $^{4}$, D. R. Alves $^{1,9}$,
T.S. Axelrod$^{1,4}$, A.C. Becker$^{2,3}$, D.P. Bennett$^{1,2,11}$, K.H. Cook$^{1,2}$, 
K.C. Freeman$^4$, K. Griest$^{2,5}$, M.J. Lehner$^5$, 
S.L. Marshall$^{1}$, D. Minniti$^{1}$,
B.A. Peterson$^4$, M.R. Pratt$^{3,6}$, P.J. Quinn$^{10}$, A.W. Rodgers$^4$, C.W.
Stubbs$^{2,3,6}$, W. Sutherland$^7$, A. B. Tomaney$^3$, T. Vandehei$^5$,
and D.L. Welch$^{8}$}
\affil { (The MACHO Collaboration) }

\altaffiltext{1}{Lawrence Livermore National Laboratory, Livermore, CA 94550\\
E-mail:  alcock, alves, kcook, dminniti, stuart@llnl.gov}

\altaffiltext{2}{Center for Particle Astrophysics, University of California,
Berkeley, CA 94720}

\altaffiltext{3}{Department of Astronomy, University of Washington,
Seattle, WA 98195 \\
E-mail: stubbs, becker, austin@astro.washington.edu}

\altaffiltext{4}{Mt.  Stromlo and Siding Spring Observatories, Australian National
University, Weston, ACT 2611, Australia \\ E-mail: robyn, kcf, peterson, 
pjq, tsa, alex@merlin.anu.edu.au}

\altaffiltext{5}{Department of Physics, University of California,
San Diego, CA 92093 \\
E-mail: griest, vandehei, matt@astrophys.ucsd.edu}

\altaffiltext{6}{Department of Physics, University of California,
Santa Barbara, CA 93106 \\
E-mail: mrp@lensing.physics.ucsb.edu}

\altaffiltext{7}{Department of Physics, University of Oxford,
Oxford OX1 3RH, U.K.\\
E-mail: wjs@oxds02.astro.ox.ac.uk}

\altaffiltext{8}{Department of Physics and Astronomy, McMaster University,
Hamilton, ON L8S 4M1, Canada\\
E-mail: welch@physics.mcmaster.ca}

\altaffiltext{9}{Department of Physics, University of California, Davis, CA 95616}

\altaffiltext{10}{European Southern Observatory, Karl Schwarzschild Str. 2, D-85748 Garching bei M\"unchen Germany\\
E-mail: pquinn@eso.org}

\altaffiltext{11}{Physics Department, University of Notre Dame, Notre Dame, IN 46556\\
E-mail: dbennett@.nd.edu}

\pagebreak
\begin{abstract}
The recent suggestion by Zhao (1996) that the microlensing
events observed towards the Large Magellanic Cloud are due to an
intervening Sgr-like dwarf galaxy is examined. 
A search for foreground RR Lyrae in the MACHO photometry database
yields 20 stars whose distance distribution follow the expected
halo density profile.  Cepheid and red giant branch clump stars
in the MACHO database are consistent with membership in the LMC.
There is also no evidence in the
literature for a distinct kinematic population, for intervening gas,
or for the turn-off of such a hypothetical galaxy.
We conclude that if the lenses are in a foreground galaxy, 
it must be a particularly dark galaxy.

\end{abstract}

\pagebreak

\section{Introduction}
The MACHO and EROS projects have detected microlensing towards the
Large Magellanic Cloud --LMC-- 
(Alcock et al. 1996, 1997, Aubourg et al. 1995), and 
interpreted these events as due to lenses located in the Milky Way halo.
There have been, however, other interpretations of the LMC 
microlensing events.

Sahu (1995) argued that the lenses may be located in the disk of the 
LMC itself.  This is unlikely because the spatial distribution of the events
is not clustered towards the LMC bar, and because this model predicts
only 1 event in the 2-yr data of Alcock et al. (1997), while 6-8 bonafide 
events are observed. We note, however, that the binary event LMC-9 appears likely
to be located in the LMC, if the LMC velocity dispersion (and 
microlensing optical depth) is small (Bennett et al. 1996).

Gould et al. (1996) argue that one of the MACHO events (LMC-5) could be due to 
an M star lens located in the Milky Way disk. This is possible, as Alcock et 
al. (1997) estimate that about one event out of the eight detected should be 
due to a thin- or thick-disk lens. 

Zhao (1996) has suggested that the LMC 
microlensing could be due to an intervening dwarf galaxy like the Sgr dwarf
(Ibata et al. 1995). The suggestion is supported by an apparent
peak in the RR Lyrae distribution of Payne-Gaposchkin (1971) at 16 to 25 kpc.
However, Connolly (1985) suggest that these faint RR Lyrae may be associated
with the LMC, and thus may not be RR Lyrae at all. These stars would have
periods similar to RR Lyrae, but absolute magnitudes that are about 2
magnitudes brighter, if they belong to the LMC. 
This issue can be readily explored using the MACHO photometric database 
and other recent observations reported in the literature. In this letter
we examine in detail
the RR Lyrae distance distribution, concluding that there 
is no evidence of the existence of such a galaxy.
More recently, Zhao (1997) proposed that a significant fraction of the
microlensing events may be caused by lenses within an extended tidal tail
in front of the LMC. This very interesting suggestion may be addressed in
the future when variable stars and microlensing towards
new lines of sight, such as the Small Magellanic Cloud,  are explored.

\section{Search for a Dwarf Galaxy in the MACHO Database}
We will look for evidence of an intervening galaxy in between us and
the LMC in the MACHO photometric database. Dwarf 
galaxies of the Local Group have experienced complex
star formation histories, containing at present old and intermediate age stellar
populations (see reviews by Hodge 1994 and Mateo 1996). 
Thus, we will search for these populations
using known tracers: RR Lyrae, Cepheid and clump giant stars. These tracers
are good distance indicators, and they would be brighter if they belong
to an intervening galaxy located in front of the LMC.
Zhao (1996) uses the specific example of a galaxy like the Sgr dwarf discovered
by Ibata et al. (1995).  This hypothetical Sgr-like dwarf galaxy would be
located between the outskirts of the
Galactic disk, and the LMC, with distance modulus
constrained to the range $11.5 < (m-M)_0 < 18.5$ mag. We also assume that the
line of sight depth of this galaxy is small, $\Delta R < 3$ kpc
(i.e. comparable to the observed depth of the Sgr dwarf, Alcock et al.
1997), a priori discarding a ``finger of God" effect.

\subsection{The RR Lyrae Stars}
The RR Lyrae stars are easily detected tracers of an intervening galaxy, because they
are easily identified from their light curves, and they are excellent distance
indicators. RR Lyrae
belonging to a dwarf galaxy that is not extended along the line of sight
will show as a peak in the luminosity function, while RR Lyrae
belonging to the Milky Way halo would be more or less uniformly 
distributed between the Sun and the LMC. For example,
the Sgr galaxy shows up clearly as an extra peak in the RR Lyrae luminosity 
function of bulge fields (Alard 1996, Mateo et al. 1996, Alcock et al. 1997).

Upon inspection of the data presented by Payne-Gaposchkin (1971),
Zhao (1996) suggests that there are a number of RR Lyrae
clustered at 16 to 25 kpc, arguing that they trace an over-dense region,
with significant consequences for the interpretation of microlensing towards 
the LMC.  The spectroscopic followup of 8 of these stars by
Connolly (1985) and Smith (1985) confirmed that 3 of them are not RR Lyrae, and
that 5 of them are galactic halo RR Lyrae stars, not members of the LMC. 
The radial velocities also prove
that these 5 stars are not  related to each other: $\sigma_{obs} = 147$ km s$^{-1}$
(for comparison, $\sigma_{Sgr} = 11$ km s$^{-1}$, Ibata et al. 1995).

We have studied the line-of-sight distribution of the RR Lyrae stars
in front of the LMC using the year-one MACHO photometry database,
which contains $5\times10^{4}$ variable stars
in the LMC bar region.
The period-amplitude diagram for $10^4$ short-period LMC variable stars is
presented in Figure 1,  showing the cuts applied
to select candidate RR Lyrae types ab and c. 

Figure 2 shows the color-magnitude diagram of RR Lyrae star candidates from 
the cuts applied in the period-amplitude plane of Figure 1.  Magnitudes are
approximate Kron-Cousins $V$ and $R$, adopting a global calibration
solution and neglecting small field to field zero point differences.
The left panel shows the vast majority of LMC RRab ($N=6700$)
centered at about $V=19.1$, $V-R=0.25$. In order to select foreground 
RR Lyrae stars we will consider only stars with $V<18$.
There is a large group 
of stars ($N\approx 200$) centered at $V=18.3$, $V-R=0.35$. These are LMC
RRab blended with clump giant stars, according to their brighter magnitudes,
redder colors and smaller amplitudes. There is an extended group of
blue stars ($N\approx 600$) with $V-R <0.1$, 
which include some RRab blended with main-sequence stars and
many eclipsing binaries.
Finally, there is a group of stars centered at about
$V=17.0$, $V-R=0.25$, which are mostly short-period type I Cepheids in the LMC
(Alcock et al. 1997, in preparation).  
They are a major contaminating source when selecting
foreground RRab stars, because they have similar periods, amplitudes and colors
to those stars.  Foreground RR Lyrae and type I Cepheids were distinguished
using the shape of their lightcurves.  We also checked the location of these stars
in the period-magnitude diagram and compared them with the MACHO type I
Cepheid sample.  The short period Cepheids clearly lie along the
1H (overtone) sequence.  Some of these Cepheids turned out to be
newly discovered 1H/2H beat Cepheids, identified via excess photometric
scatter at maximum brightness and confirmed by discrete Fourier transforms
(Alcock et al. 1995). 
The right panel of Figure 2 shows the RR Lyrae type c stars ($N=1800$), with
about $N\approx 200$ eclipsing binaries and RRc stars blended with
clump giants and main sequence stars.   

We find 16 foreground RRab stars with amplitudes $A>0.25$, periods $0.4<P<0.7$ days, 
colors $0.2<V-R<0.35$, and magnitudes
$V<18$, and 4 foreground RRc stars with $A>0.25$, $0.2<P<0.4$ days, $0.1<V-R<0.25$, and
$V<18$.  These cuts are optimized to discriminate foreground RR Lyrae from
LMC RR Lyrae, blends and other variable stars.  The stars were additionally
identified in the four-year MACHO database.  Calibrated
lightcurves confirm their identification as RR Lyrae and  do not change
the distance distribution.  The coordinates, magnitudes, colors and periods
of the present sample are listed in Table 1.

We measure distances to the foreground RR Lyrae stars adopting $M_V = +0.4$
(Reid 1997, Gratton et al. 1997). 
The uncertainty in these individual distances is estimated to be about 10\%,
with the caveat that the zero point in RR Lyrae absolute magnitudes
may be more uncertain (e.g. $M_V > 0.53$ Gould 1995), which does not have
any effect in the results of this paper.
The distance distribution of the 20 foreground RR Lyrae stars in the direction of 
the LMC is shown in Figure 3. The lines illustrate a 
$r^{-3.5}$ density halo
with spherical flattened (b/a=0.6) distributions.
The flattened halo is normalized arbitrarily to give the same total
number of stars out to 40 kpc. 
Note that there is no significant concentration at any distance. Furthermore,
the distribution of these RR Lyrae stars in the sky is uniform across the LMC,
not concentrated or clumped. 

The observation of RR Lyrae stars in the Sgr dwarf by Alcock et al. (1997)
allows us to give a direct comparison with what would be expected if
there were a dwarf galaxy in the foreground. We place this set of RRab stars
($N=34$) in the direction of the LMC, given that the total
areas surveyed towards the LMC and bulge are similar (neglecting incompleteness
corrections, which would be worse in the bulge fields, and neglecting RRc stars
in Sgr, which would increase this sample to $N\sim 50$, but which are incomplete).
The distance distribution of the Sgr RRab, equivalent to about $0.5$\% of all
the MACHO RRab found in the LMC, is overplotted in 
Figure 3 (we have actually used the reddening independent magnitudes
to compute their individual distances
due to the heavy differential reddening of the bulge fields). This magnitude
distribution is very different from that of the observed foreground RR Lyrae 
stars, arguing against the existence of a
Sgr-like galaxy in front of the LMC. More stringent limits are possible
once radial velocities can be obtained for the stars in our sample.

\subsection{Other Possible Evidence}
RR Lyrae trace old and metal-poor stellar populations. Classical Cepheid
variable stars arise from a different parent population, with
intermediate masses and ages.
There is no evidence for a population of foreground objects
among the 1500 Cepheids found in the inner 22 fields of the
LMC (Alcock et al. 1995). Note that every Cepheid would be 
accompanied by about 100 times their number in main-sequence
B-type stars, which are not present.

RR Lyrae and Cepheid stars represent rare stages of stellar evolution.  It may
then be possible that only a few of them belonging to an intervening dwarf
galaxy would be present in the fields explored here.   Core He-burning
clump giants should be much more numerous, since stellar evolution theory
predicts that many low mass stars go through this phase of
evolution.  

We have assembled
MACHO project photometry, properly calibrated to $V$ and $R_{KC}$, for
9 million stars in the LMC bar region into a composite
color-magnitude diagram.
The large number of
stars in this diagram reveal
many low level features tracing
short-lived stages of stellar evolution.  
The region of the color-magnitude diagram near the red giant branch
and locus of clump giants is
shown in Figure 4.
The clump giants peak near
$V-R = 0.45$ mag and $R = 18.6$ mag.  There is an extra ``bump''
seen near $V-R = 0.55$ mag and $R = 17.7$ mag.
Zaritsky (1997) sees a similar feature in the LMC.
The color of the
bump is inconsistent with these stars being clump giants
from the LMC or SMC residing in a tidal tail or stream.
If the bump stars are
clump giants of a dwarf galaxy in front of the LMC,
they are likely of higher
metallicity than the LMC clump giants, since reddening 
cannot account for their color.  While some of these bump stars are
likely clump-clump blends,  the naturally occurring
stellar populations of the LMC are the most plausible 
explanation for this feature.
A detailed analysis
of the 9 million star color-magnitude diagram using LMC
cluster photometry, theoretical isochrones, and
MACHO variable star data will be presented elsewhere
(Alves, 1998).

It is very difficult to detect overlapping nearby galaxies, as proven by the
recent discovery of the Sgr dwarf behind the Milky Way bulge. 
Effective ways of searching for such systems would be distance indicators
(RR Lyrae stars, Cepheid stars, clump giants, discussed above), radial 
velocity surveys, and star counts.
Extensive star counts and surface brightness work
by de Vaucouleurs (1955), and more recently by Bothun \& Thompson
(1988) show no evidence for anything that is not LMC-centric.

If there were such a galaxy,
we should be able to see a distinctive old main-sequence turn-off in deep
HST photometry. This is because main sequence turn-off stars are 
much more numerous than the tracers discussed above. A clear example 
of this is the detection of the main sequence turn-off of the Sgr galaxy 
behind the bulge by Fahlman et al. (1996) using WFPC2 photometry.  
Olszewski et al. (1996) review the recent color-magnitude diagrams of the 
LMC.  We note that none of the recent deep HST color-magnitude diagrams of the LMC 
(e.g. Elson et al. 1994, Holtzman et al 1997) show
such an extra stellar component.

Different authors have published radial velocities of giant stars
in LMC fields (see Olszewski et al. 1991).
No evidence is found for grouping in velocity space of stars that
may belong to a dwarf galaxy in front of the LMC.

Also, the Sgr dwarf has four globular clusters sharing the same kinematics
and spatial distribution (DaCosta \& Armandroff 1995). However,
the old globular clusters located in the direction of the
LMC have kinematics and distances consistent with 
membership to this galaxy (Suntzeff et al. 1992), and
there is no evidence for clusters belonging to an 
unknown Sgr-like dwarf galaxy.

The presence of a gas rich dwarf galaxy (dIrr) is also ruled out
because we do not see large amounts of gas in front of the LMC. 
Extensive searches have only revealed HI detection belonging to gas 
in the Magellanic stream (Rohlfs et al. 1984, Heller \& Rohlfs 1994).
Even though there is evidence for interstellar lines at many velocities in 
high resolution spectra of SN~1987A (Savage et al. 1989), 
absorption line studies of LMC stars do not reveal an significant
gaseous component other than that of the Milky Way or the 
LMC disk (Westerlund 1990, Roth \& Blades 1997).
To our knowledge, there is no other evidence reported in the literature 
regarding a galaxy in front of the LMC. 

\section{Summary}
We have tested the idea of Zhao (1996) that the microlensing events observed 
in LMC fields may be due to an intervening dwarf galaxy like the Sgr dwarf.
Searching for foreground RR Lyrae in the year-one MACHO photometry database
yielded 20 stars whose distribution follows the expected halo density law.
These foreground RR Lyrae represent 0.3\% of the total number of LMC RR Lyrae
stars with similar colors detected in these fields.
Classical Cepheid variable stars and clump giants seen in the MACHO
database are consistent with membership in the LMC.
We do not find any evidence for the stellar 
component of such a dwarf galaxy in the MACHO database,
from which we conclude that if the lenses are indeed in a foreground galaxy, this
must be a particularly dark galaxy.  Alternatively, some of the microlensing
events may be due to an extended tidal tail (Zhao 1997). This is more difficult to
test with the present data, but could be addressed when other lines of
sight (such as the SMC) are studied.

\acknowledgements
{
We are also very grateful for the skilled support given to our project 
by the technical staff at the Mt. Stromlo Observatory.  
Work performed at LLNL is supported by the DOE under contract W7405-ENG-48.
Work performed by the Center for Particle Astrophysics on the UC campuses
is supported in part by the Office of Science and Technology Centers of
NSF under cooperative agreement AST-8809616.
Work performed at MSSSO is supported by the Bilateral Science 
and Technology Program of the Australian Department of Industry, Technology
and Regional Development.   WJS is supported by a PPARC Advanced Fellowship. 
KG  and MJL acknowledge support from DoE, Alfred P. Sloan, and Cottrell awards.
CWS thanks the Sloan, Packard and Seaver Foundations for their support.
DM thanks the Aspen Center for Physics for their kind hospitality, as well as
useful comments and suggestions from A. Gould, H.-S. Zhao, and D. Zaritsky.
}

\begin{figure}
\plotone{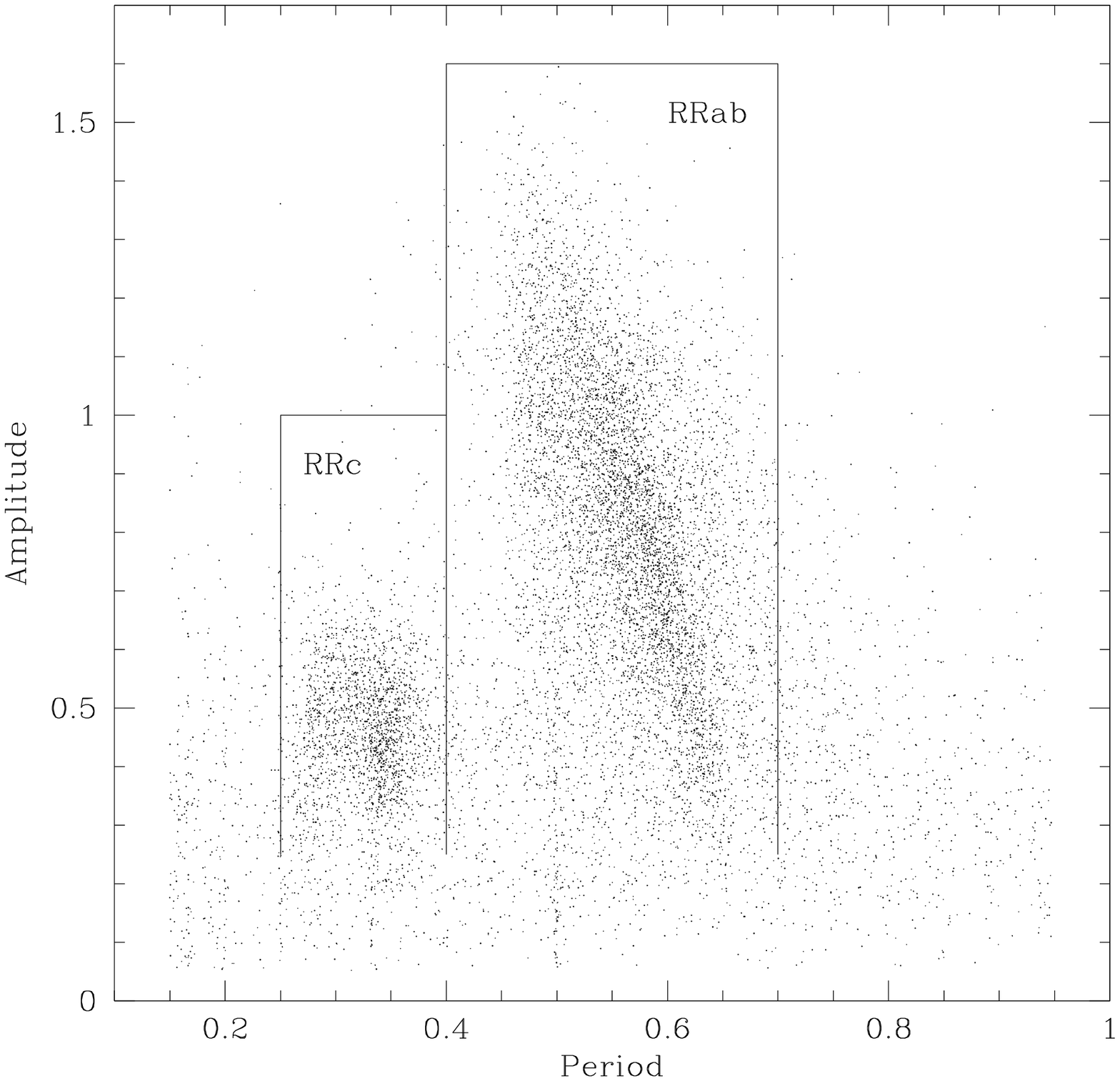}
\caption{ 
Period-amplitude diagram from the MACHO Project $V$ light-curves showing the cuts to 
select candidate RR Lyrae types ab and c. Note the concentration of a few 
alias periods at 0.333 and 0.5 days.
}
\end{figure}

\begin{figure}
\plotone{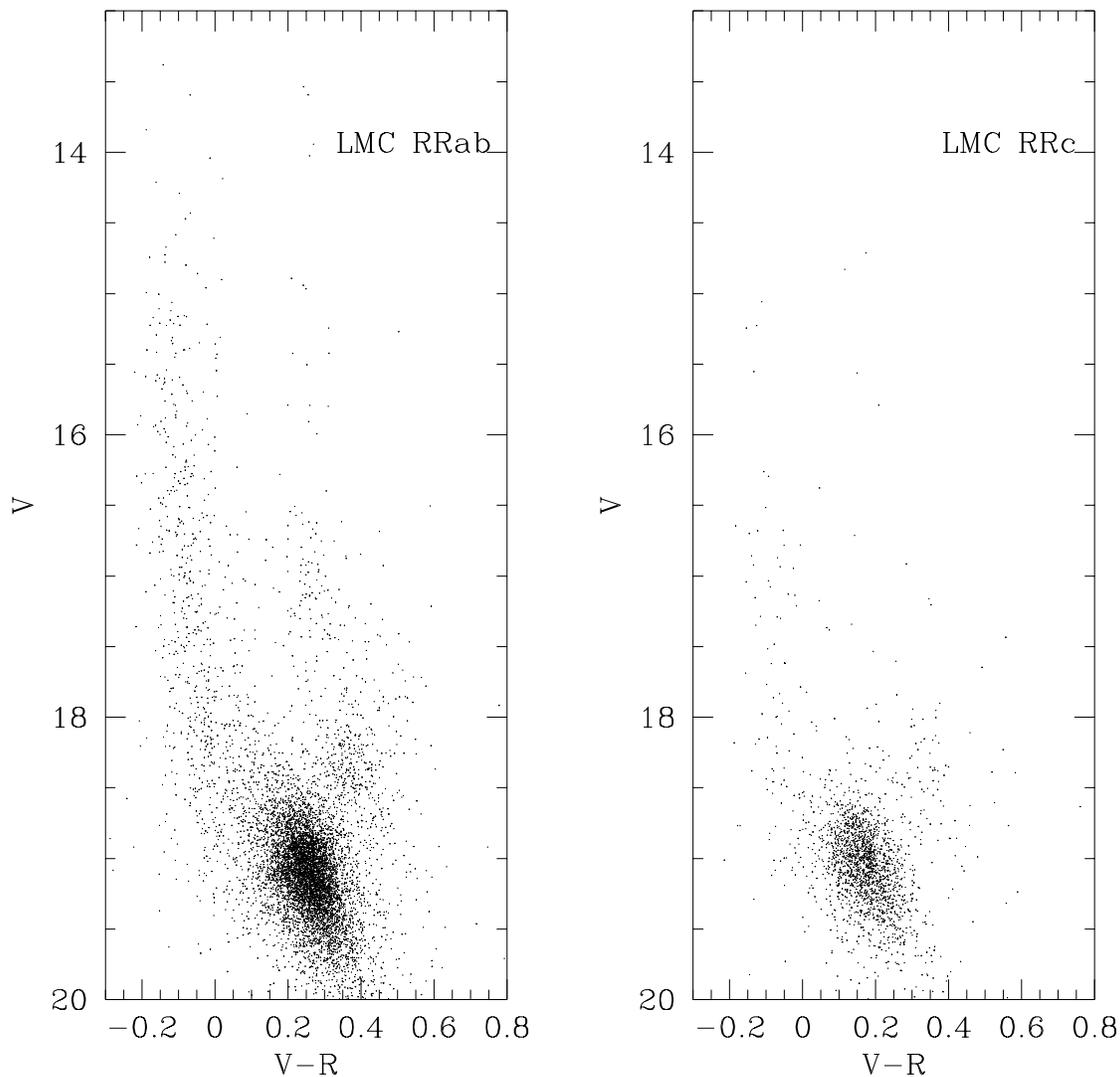}
\caption{ 
Color-magnitude diagram showing RR Lyrae star candidates from the cuts
applied in the period-amplitude plane (c.f. Figure 1). The majority of the
LMC RR Lyrae are seen at $V=19.1$, with colors and magnitudes apread
along the reddening vector.  Note that the
concentration of stars redder and brighter than the LMC RR Lyrae are
RR Lyr-clump giant blends, that the bluer sequence of brighter stars are
RR Lyr-main sequence blends, and that the group of stars located about 
2 magnitudes above the RR Lyrae type ab are mostly short period Cepheids.
}
\end{figure}

\begin{figure}
\plotone{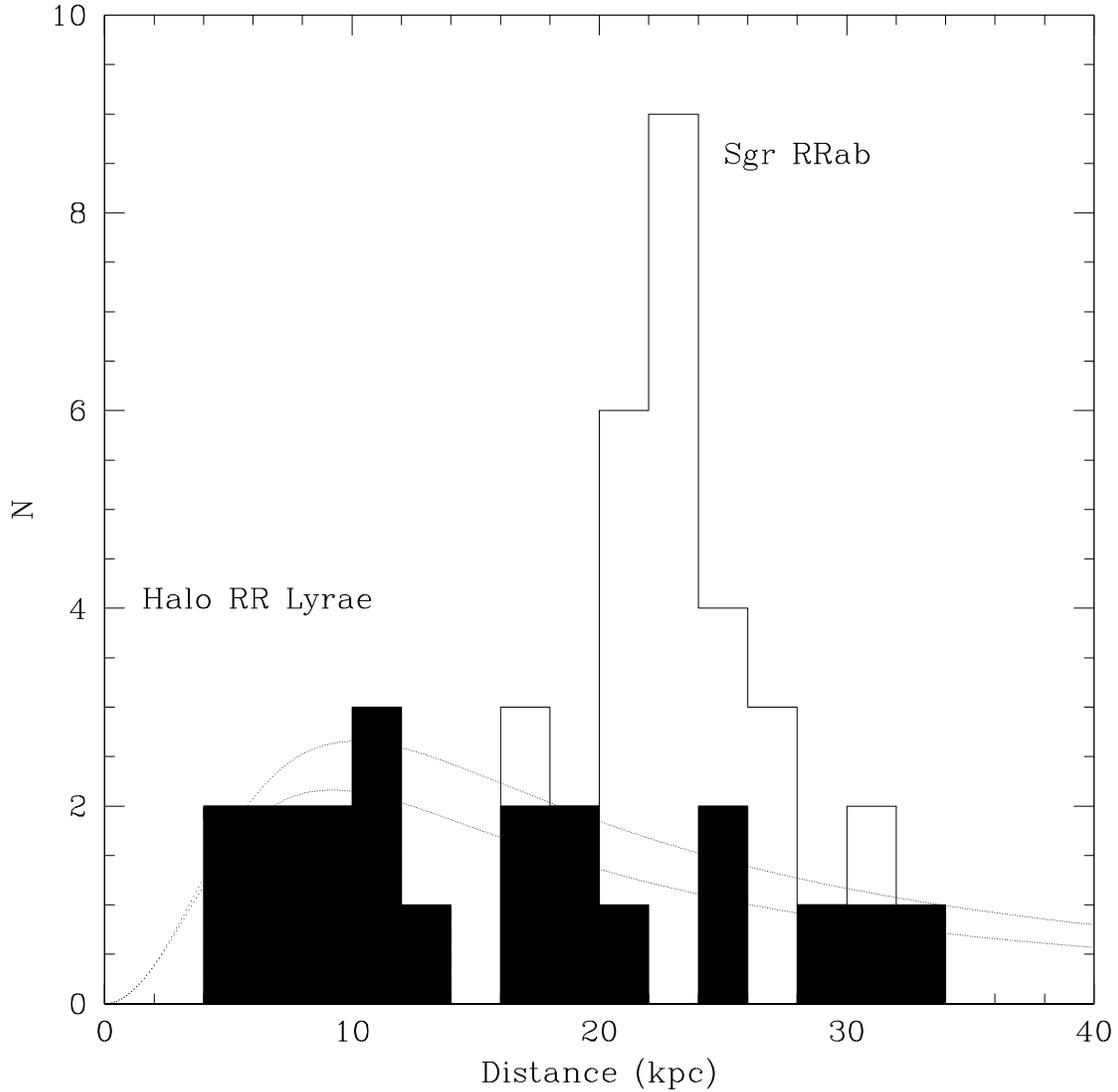}
\caption{ 
Distance distribution of foreground RR Lyrae stars in the direction of the LMC
(solid histogram).  The upper and lower dotted lines indicate a $r^{-3.5}$ density halo
with spherical and flattened (b/a=0.6) distributions, respectively.
Note that there is no significant concentration at any distance.
The distance distribution of Sgr
RRab observed in bulge fields (Alcock et al. 1997) is overplotted
(open histogram) as an example of what we should have seen if there were
a foreground Sgr-like galaxy.
}
\end{figure}

\begin{figure}
\vskip -2truecm
\plotone{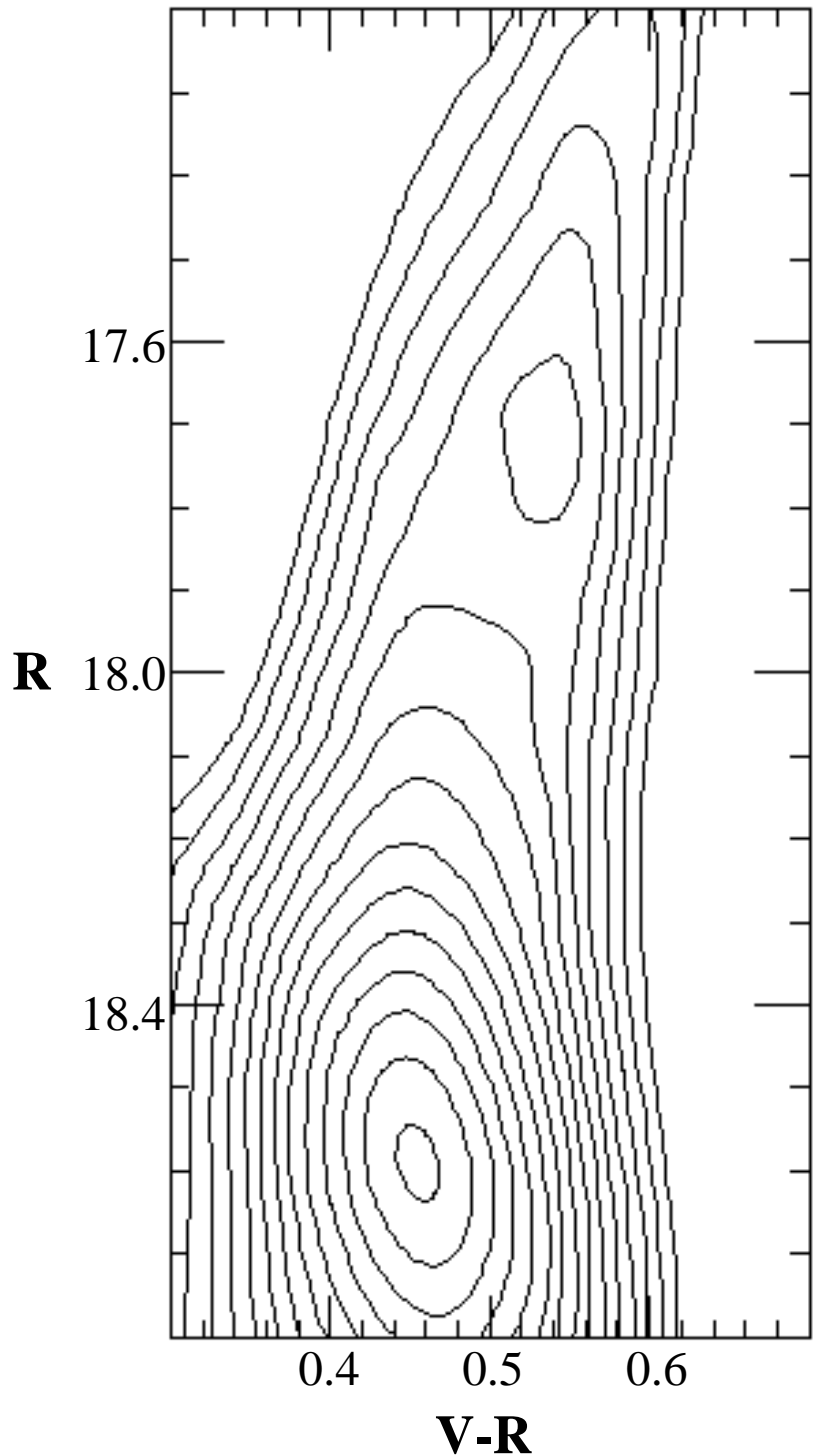}
\caption{The clump giant region of the
9 million star color-magnitude diagram of the LMC bar.  Pixel size
is 0.04 mag in $V-R$ and 0.08 mag in R.  Contours represent the number
of stars in each color-magnitude diagram pixel.  Contours are spaced by 0.1 dex,
the lowest contour is 3.5 dex and the highest contour is 5.0 dex,
with the pixel size being 0.04 and 0.08 mag in $V-R$
and $R$, respectively.  This figure shows the well populated clump near $V-R = 0.45$ mag
and $R = 18.6$ mag.  Superimposed on the red giant branch near
$V-R = 0.55$ mag and $R = 17.7$ mag is the bump discussed in the text.
}
\end{figure}

\begin{deluxetable}{lrllllllll}
\small
\footnotesize
\tablewidth{0pt}
\scriptsize
\tablecaption{Foreground RR Lyrae Stars Towards the LMC}
\tablehead{
\multicolumn{1}{c}{Macho ID}&
\multicolumn{1}{c}{P (day)}&
\multicolumn{1}{c}{$\langle V \rangle$}&
\multicolumn{1}{c}{$\langle R \rangle$}&
\multicolumn{1}{c}{Type}&
\multicolumn{1}{c}{$RA_{2000}$}&
\multicolumn{1}{c}{$DEC_{2000}$}&
\multicolumn{1}{c}{Comments}}
\startdata
1.4297.33	&0.68819&	15.05&	14.86&	ab&	05:06:50.396&	-68:49:38.06&\\
15.10307.80	&0.51864&	17.64&	17.29&	ab&	05:44:18.167&	-71:26:15.89&\\
19.4302.382	&0.55915&	16.20&	15.99&	ab&	05:06:48.759&	-68:26:40.18&\\
6.5722.3	&0.55342&	13.73&	13.55&	ab&	05:15:50.055&	-70:36:46.56&	sat. at max, also 13.5722.2833\\
11.9107.197	&0.60315&	18.13&	17.79&	ab&	05:36:13.510&	-70:48:44.40&\\
5.5125.3195	&0.65617&	18.23&	17.93&	ab&	05:11:51.262&	-70:01:43.98&\\
5.5495.70	&0.61461&	16.61&	16.36&	ab&	05:14:33.476&	-69:37:06.10&\\
81.9001.2087	&0.55881&	17.17&	16.97&	ab&	05:35:55.304&	-69:48:27.85&\\
82.8410.55	&0.51445&	16.27&	16.08&	ab&	05:32:18.963&	-68:51:52.15&	Blazkho\\
82.7928.95	&0.47585&	17.06&	16.88&	ab&	05:29:08.602&	-68:42:34.57&\\
10.3187.111	&0.54498&	17.78&	17.44&	ab&	05:00:09.002&	-70:09:59.20&\\
14.8374.25	&0.62844&	15.70&	15.48&	ab&	05:32:11.964&	-71:13:50.90&\\
18.2598.10	&0.63913&	15.21&	14.95&	ab&	04:56:29.268&	-69:08:25.32&	Blazkho?\\
18.3208.40	&0.66432&	15.83&	15.64&	ab&	05:00:21.175&	-68:46:21.00&\\
19.3825.18	&0.58675&	15.62&	15.42&	ab&	05:04:09.837& 	-68:00:21.30&\\
80.7071.3069	&0.55312&	13.88&	13.67&	ab&	05:24:06.126&	-69:25:11.93&also	77.7071.12\\
3.6718.38	&0.33555&	15.89&	15.72&	c&	05:21:53.228&	-68:42:37.02&also 	80.6718.4092\\
14.8371.68	&0.32683&	16.93&	16.79&	c&	05:32:24.384&	-71:29:03.81&\\
18.2485.17	&0.30745&	14.88&	14.75&	c&	04:55:29.721&	-68:35:36.87&\\
5.4644.8	&0.32799&	15.14&	15.04&	c&	05:09:18.712&	-69:50:15.31&	period changes\\
\enddata
\end {deluxetable}


\begin{references}
\reference{}{Alard, C. 1996, ApJ, 458, L17}
\reference{}{Alcock et al. 1995, AJ, 109, 1653}
\reference{}{Alcock et al. 1996, ApJ, 461, 84}
\reference{}{Alcock et al. 1997, ApJ, 474, 217}
\reference{}{Alves, D. R. 1998, Ph.D. dissertation, University of California at Davis, in preparation}
\reference{}{Aubourg, E., et al. 1995, A\&A, 301, 1}
\reference{}{Bennett, D., et al. 1996, Nucl.Phys.Proc.Suppl., 51B, 152}
\reference{}{Bothun, G. D., \& Thompson, I. B. 1988, AJ, 96, 877}
\reference{}{Connolly, L. P. 1985, ApJ, 299, 728}
\reference{}{Elson, R. W., Forbes, D. A., \& Gilmore, G. F. 1994, PASP, 106, 632}
\reference{}{Fahlman, G. G., Mandushev, G., Richer, H. B., Thompson, H. B., \& Sivaramakrishnan, A. 1996, ApJ, 459, L65}
\reference{}{Gould, A., Bahcall, J. N., \& Flynn, C. 1996, ApJ, 465, 759}
\reference{}{Gould, A. 1995, ApJ, 452, 189}
\reference{}{Gratton, R., Fusi Pecci, F., Carretta, E., Clementini, G., Corsi, C. E., Lattanzi, M. 1997, ApJ, in press (astro-ph/9704150)}
\reference{}{Heller, P., \& Rohlfs, K. 1994, A\&A, 291, 743}
\reference{}{Hodge, P. 1994, in "The Local Group: Comparative and Global Properties", eds. A. Layden, R. C. Smith, \& J. Storm (ESO: Garching), p. 57}
\reference{}{Holtzman, J., et al. 1997, AJ, 113, 656}
\reference{}{Ibata, R., Gilmore, G., \& Irwin, M. 1995, MNRAS, 277, 781}
\reference{}{Mateo, M. 1996, in ASP Conf. 92 on "Formation of the Galactic Halo... Inside and Out", eds. H. Morison \& A. Sarajedini, (ASP: San Francisco), p. 434.}
\reference{}{Mateo, M., Mirabal, N., Udalski, A., Szymanski, M., Kaluzni, J., Kubiak, M., Krzeminski, W., \& Stanek, K. Z. 1996, ApJ, 458, L13}
\reference{}{Olszewski, E. W., Schommer, R. A., Suntzeff, N. B., \& Harris, H. C. 1991, AJ, 101, 515}
\reference{}{Olszewski, E. W., Suntzeff, N. B., \& Mateo, M. 1996, ARA\&A, 34, 511}
\reference{}{Payne-Gaposchkin, C. 1971, Smithsonian Contr. Ap. No. 13}
\reference{}{Rohlfs, K., Kreitschann, J., Siegman, B. C., \& Feitzinger, J. V. 1984, A\&A, 137, 343}
\reference{}{Roth, K. C., \& Blades, J. C. 1997, ApJ, 474, L95}
\reference{}{Reid, I. N, 1997, AJ in press (astro-ph/9704078)}
\reference{}{Sahu, K. C. 1994, Nature, 370, 275}
\reference{}{Smith, H. A. 1985, PASP, 97, 1053}
\reference{}{Savage, B. D., Jenkins, E. B., Joseph, C. L, \& De Boer, K. S. 1989, ApJ, 345, 393}
\reference{}{Suntzeff, N. B., Schommer, R. A., Olszewski, E. W., \& Walker, A. R. 1992, AJ, 104, 1743}
\reference{}{de Vaucouleurs, G. 1955, AJ, 60, 126}
\reference{}{Westerlund, B. E. 1990, A\&A Rev., 2, 29}
\reference{}{Zaritsky, D. 1997, private communication}
\reference{}{Zhao, H.-S. 1996, MNRAS submitted (astro-ph/9606166)}
\reference{}{Zhao, H.-S. 1997, MNRAS submitted (astro-ph/9703097)}
\end{references}
\end{document}